\documentstyle[12pt,epsf]{article}
\def\be{\begin{eqnarray}}
\def\ee{\end{eqnarray}}
\begin{document}
\vspace*{1cm}
\begin{flushright}
hep-ph/9410332\\
TTP 94-21\\
\end{flushright}

\vspace*{2cm}
\noindent
{\bf  ON THE NEW METHOD OF COMPUTING\\ TWO-LOOP MASSIVE DIAGRAMS
\footnote{Submitted to the proceedings of
the NATO Advanced Study Institute ``Frontiers in Particle
Physics'', Carg\`ese, August 1994.}
}\\

\vspace{1.2cm}
\hspace*{19mm}
\parbox{4in}{
\noindent
Andrzej Czarnecki

\vspace{5mm}
\noindent
Institut f\"ur Theoretische Teilchenphysik\\
 Universit\"at Karlsruhe\\
 D-76128 Karlsruhe, Germany\\
e-mail: ac@ttpux2.physik.uni-karlsruhe.de}\\
\vspace{1.2cm}



\noindent
{\bf INTRODUCTION}\\

The improving precision of experiments in the high energy physics
motivates theoretical studies of quantum corrections to various
processes.  In the two-loop approximation this is connected with
great computational difficulties, especially if there are several mass
scales involved in the process, which is a typical situation in the
case of electroweak or mixed chromodynamic and electroweak
corrections.  Recently a new method has been proposed for the
evaluation of scalar two-loop vertex and propagator functions
\cite{vertexd,ckk}.  It has also been shown that a similar approach
works even for the four-point functions \cite{shnote}. In this talk I
present a few examples which illustrate the principle of this method.

The aim will be to obtain a double integral representation which is
suitable for numerical evaluation.  In the following section I will
derive it for a special case of the vertex function with zero momentum
transfer.  The next section refers to the general case of a planar
vertex function with space-like values of external momenta, and the
last one shows an example of dealing with ultraviolet divergent
diagrams. The examples of two-loop functions to be considered in this
paper are depicted in Fig.~1(a,b,c).
\\
\vspace{1cm}

\noindent
{\bf VERTEX FUNCTION AT ZERO MOMENTUM TRANSFER}\\

While ref.~\cite{ckk} describes the general method of computing the
two-loop vertex function, here the method is illustrated with the
special case of zero-momentum transfer which is in fact a two-point
function.
The principle remains the same, but the
computation becomes much simpler and it is easy to write down explicit
formulas.
The diagram and numbering of lines is depicted in
Fig.~1(b). The four momenta in the rest frame of the external
particle are ($p_i$ are outgoing)
\be
p_2^\mu&=&-p_1^\mu=(q,0,0,0),\nonumber\\
l^\mu&=&(l_++l_-,l_+-l_-,\vec l_\bot),\nonumber\\
k^\mu&=&(k_++k_-,k_+-k_-,\vec k_\bot),
\ee
and the momentum $k$ runs along the lines (4,3,5,6) and $l$
along (1,3,2).  We shall see later that one obtains different,
but equivalent, formulas if one uses a different routing of the momenta.
The two-loop function of Fig.~1(b) is
\be
V_0(q^2)=\int \!\! \int d^4k d^4l {1\over P_1P_2P_3P_4P_5P_6}.
\label{eq:vi}
\ee
With the present choice of momenta, and with $s\equiv l_\bot^2$ and
$t\equiv k_\bot^2$,
the explicit form of the propagators is
\be
P_{1,2}&=&(2l_+-q)(2l_--q)-s-m_{1,2}^2+i\eta,\nonumber\\
P_3&=&4(l_++k_+)(l_-+k_-)-s-t-2\sqrt{st}z-m_3^2+i\eta,\nonumber\\
P_{4,5}&=&(2k_++q)(2k_-+q)-t-m_{4,5}^2+i\eta,\nonumber\\
P_6&=&4k_+k_--t-m_6^2+i\eta.
\label{props}
\ee
There is only one propagator, $P_3$, through which both internal
momenta flow, and $z$ denotes the cosine of the angle between the two
perpendicular momentum vectors $\vec k_\bot$ and $\vec l_\bot$.
Integrations over the two angular variables describing the absolute and
relative configuration of $\vec k_\bot$ and $\vec l_\bot$ can now be
performed and we obtain from (\ref{eq:vi})
\be
V_0(q^2)=-4\pi^2\int dk_+ dl_+ds dtdk_- dl_-
 {1\over \sqrt{A^2-B^2}} {1\over P_1P_2P_4P_5P_6},
\label{angul}
\ee
with
\be
A&=&4(l_++k_+)(l_-+k_-)-s-t-m_3^2+i\eta,\nonumber\\
B^2&=&4st.
\ee
The integrations over the $k_-$ and $l_-$ are done with help of
contour integrals. It turns out that the singularities in $P_4$ and
$P_5$ do not contribute, while $P_1$, $P_2$ and $P_6$ contribute only for
$k_+$ and $l_+$ lying in a triangular region $T$ in the $k_+l_+$ plane
 delimited by the lines
\be
k_++l_+&=&0,\nonumber\\
k_+&=&0,\nonumber\\
l_+&=&{q\over 2}.
\label{tri}
\ee
The function $V_0$ becomes now
\be
\lefteqn{V_0(q^2)=2\pi^4\int\!\!\int_{T}{dk_+ dl_+ \over k_+(2l_+-q)}
\int_0^\infty dt \int_0^\infty
ds}
\nonumber\\
&& \times
\sum_{\{i,j\}=\{1,2\},\{2,1\}} \left.{1\over P_iP_4P_5}{1\over
\sqrt{A^2-B^2}} \right|_{k_-\to k_6,
l_-\to l_j},
\ee
where
\be
k_6={t+m_6^2\over 4k_+},\qquad l_{1,2}={s+m_{1,2}^2\over 2(2l_+-q)}+{q\over 2}.
\ee
Substituting the explicit formulas for the propagators one obtains
\be
\lefteqn{V_0(q^2)={8\pi^4\over q^2(m_2^2-m_1^2)}
\int\!\!\int_{T}dk_+ dl_+{k_+ \over 2l_+-q}
\int_0^\infty {dt\over (t+t_4)(t+t_5)}} \nonumber\\
&&\times\int_0^\infty
ds
\left({1\over \sqrt{(at+b_2+cs)^2-4st}} -{1\over
\sqrt{(at+b_1+cs)^2-4st}} \right),
\ee
with
\be
a&=&{l_+\over k_+} ,\nonumber\\
b_{1,2}&=&2(k_++l_+)\left({m_6^2\over 2k_+}+{m_{1,2}^2\over
2l_+-q}+q\right)-m_3^2, \nonumber\\
c&=&{2k_++q\over 2l_+-q}, \nonumber\\
t_{4,5}
&=&{2k_+\over q}\left\{(2k_++q)\left({m_6^2\over
2k_+}+q\right)-m_{4,5}^2
\right\}.
\ee
For the sake of simplicity let us assume that $q^2$ lies below all
thresholds so that the function $V_0$ is real.
In such case the integration over $s$ is elementary
\be
\lefteqn{
\int_0^\infty ds \left({1\over \sqrt{(at+b_2+cs)^2-4st}} -{1\over
\sqrt{(at+b_1+cs)^2-4st}} \right)}\nonumber\\
& &\hspace*{6cm} = {1\over c}\ln{t(1-ca)-cb_2\over t(1-ca)-cb_1},
\ee
and in the integration over $t$ one encounters dilogarithms,
\be
{\rm Li}_2(x) = -\int_0^x {\rm d}y\, {\ln |1-y|\over  y}.
\ee
The final result is
\be
\lefteqn{V_0(q^2)={4\pi^4\over q(m_2^2-m_1^2)(m_5^2-m_4^2)}
\int_{-q/2}^0 {dk_+  \over 2k_++q}\int_{-k_+}^{q/2}dl_+}
\nonumber\\
&&\times\left\{\ln{t_4\over t_5}\ln{t_2\over t_1}
+{\rm Li}_2\left(1-{t_5\over t_2}\right)
-{\rm Li}_2\left(1-{t_4\over t_2}\right)
-{\rm Li}_2\left(1-{t_5\over t_1}\right)
+{\rm Li}_2\left(1-{t_4\over t_1}\right)\right\},\nonumber\\
\label{vi1}
\ee
with $t_{1,2}=-cb_{1,2}/(1-ca)$.

The complication which arises in the case of the non-zero momentum
transfer consists in the fact that the propagators $P_{1,2}$ after
substitution of the appropriate value for $l_-$ do not have the simple
form $\pm(m_2^2-m_1^2)$ but retain dependence on the variables $l_+$
and $s$.  This leads to a more complicated form for the integrations
over $s$ and $t$, and the final formula contains dilogarithms as well
as Clausen functions.  There are also two additional residues which
contribute and each contribution in general comes from a different
triangle in the $k_+ l_+$ plane.

Since the final two integrations over $k_+$ and $l_+$ are to be done
numerically it is useful to have an alternative formula which
provides a cross check and a test of accuracy. Such formula can be
derived by choosing the internal momenta in such way that $k$ runs
through the lines $(1,2,5,6,4)$ and $l$ through $(1,2,3)$.

With this choice the propagators are
\be
P_{1,2}&=&(2k_++2l_++q)(2k_-+2l_-+q)-s-t-2\sqrt{st}z-m_{1,2}^2+i\eta,
\nonumber\\
P_3&=&4l_+l_--s-m_3^2+i\eta,\nonumber\\
P_{4,5}&=&(2k_++q)(2k_-+q)-t-m_{4,5}^2+i\eta,\nonumber\\
P_6&=&4k_+k_--t-m_6^2+i\eta.
\ee
There are now two propagators which depend on $z$: $P_1$ and $P_2$,
but only one combination of propagators ($P_3$ and $P_6$) whose
singularities
contribute to the contour integrations over $k_-$ and $l_-$.  The triangular
region of the integration over $k_+$ and $l_+$ is now limited by
the lines
\be
k_++l_+ &=& -{q\over 2}, \nonumber \\
l_+&=&0, \nonumber\\
k_+&=&0.
\ee
After the integration over the angular variables and over $k_-$ and
$l_-$ we get
\be
\lefteqn{V_i(q^2)={\pi^4\over m_2^2-m_1^2}\int {dk_+dl_+\over
k_+l_+}\int_0^\infty {dt\over P_4P_5}}
\nonumber\\& &
\times\int_0^\infty ds \left.\left( {1\over \sqrt{A_1^{\prime 2}-B^2}}
-{1\over \sqrt{A_2^{\prime 2}-B^2}}\right)\right|_{k_-\to k_6,l_-\to
l_3}
\ee
with
\be
k_6\!\!&=&\!\!{t+m_6^2\over 4k_+}, \qquad l_3={s+m_3^2\over 4l_+},\nonumber\\
A^\prime_{1,2}&=&(2k_++2l_++q)\left(
{t+m_6^2\over 2k_+}+{s+m_3^2\over 2l_+}+q\right)-s-t -m_{1,2}^2.
\ee
The integrations over $s$ and $t$ proceed in exactly the same way as
in the previous calculation.  Finally, we can make the shift $l_+\to l_+-q/2$.
It turns out that this change of variables not only makes
the region of integration in the $k_+l_+$ plane equal to the
triangle $T$ defined in (\ref{tri}), but the whole formula for $V_i$
becomes almost the same as formula (\ref{vi1}), the only difference
being  the coefficients $b_i$, which in the present case are
\be
b_{1,2}^\prime =2(k_++l_+)\left({m_6^2\over 2k_+}+{m_{3}^2\over
2l_+-q}+q\right)-m_{1,2}^2.
\ee
The equivalence of the two formulas can be checked after
integrating over $k_+$ and $l_+$.  It provides an excellent cross check
for the numerical calculation.

In practical calculations one can encounter a mass configuration in
which $m_1=m_2$. In this case the formula simplifies:
\be
\lefteqn{V_0(q^2, m_1=m_2)=-{4\pi^4\over q^2(m_5^2-m_4^2) }}
\nonumber \\ && \times
\int_{-q/2}^0 dk_+  \int_{-k_+}^{q/2}dl_+
{k_+\over k_++l_+}
\left( {1\over t_5-t_0}\ln{t_5\over t_0}-{1\over t_4-t_0}\ln{t_4\over
t_0}
\right),
\ee
where $t_0=-cb_1^\prime/(1-ca)$.
\\
\vspace{1cm}

\noindent
{\bf SPACE-LIKE EXTERNAL MOMENTA}\\

If the external momenta have space-like values the computation of the
propagator and vertex diagrams is greatly simplified since the
internal particles do not become on-shell.
 In particular we can easily check the analytical results obtained for
the two-point and the planar three-point functions when all internal
particles are massless.  While the result for the two-point function
(see Fig.~1(a)) has been know for long time
\cite{rosner67,che81},
the much more complex formulas for the vertex functions (of both
planar and crossed topologies) have been obtained only very recently
\cite{udyad}. We present here numerical evaluation of
the vertex function with all internal masses equal $m$ and space-like
external momenta.  In the limit $m\to 0$ we reproduce the result of
\cite{udyad}.

For the numerical calculation it is convenient to choose such
reference frame that the external outgoing
momenta become (according to the notation of Fig.~1(b))
\be
p_1^\mu&=& (e,q_1,0,0),\nonumber\\
p_2^\mu&=& (-e,q_2,0,0),\nonumber\\
p_3^\mu&=& (0,-q_1-q_2,0,0).
\ee
Repeating the calculations described in the previous section we arrive
at a double integral representation which is easy to evalulate
numerically.  Fig.~2 shows  the ratio of the vertex function
\be
V(p_1^2,p_2^2,p_3^2,m^2)=\int\!\!\int {d^4kd^4l\over
P_1P_2P_3P_4P_5P_6}
\label{vert}
\ee
to the value of the vertex at zero internal masses
\be
U(p_1^2,p_2^2,p_3^2)=\left( {i\pi^2\over p_3^2} \right)^2
\Phi^{(2)}\left( {p_1^2\over p_3^2},{p_2^2\over p_3^2} \right),
\ee
where the function $\Phi^{(2)}$ has been derived in \cite{udyad}.
In formula (\ref{vert})  $P_i$ denote propagators defined analogously
to the formula (\ref{props})). For the purpose of numerical
calculation we choose one arbitrary configuration of
external momenta
$p_1^2=-1$, $p_2^2=-4$, $p_3^2=-25$.
We see that for very small masses the ratio of the two formulas
becomes unity which confirms the analytical result of Ussyukina and
Davydychev.
\\
\vspace{1cm}

\noindent
{\bf DIVERGENT INTEGRALS}\\

The method of calculation of two-loop diagrams described here is
limited to the four-dimensional space. In dealing with divergent
integrals we first have to find another diagram with the same
divergent part but simple enough to be computed analytically.  The
difference of the two diagrams can then be calculated in four
dimensions and the final result is obtained by adding the analytical
formula for the simpler diagram. Such procedure, based on the
representation of the two-point functions proposed in~\cite{propagd},
has been described in~\cite{inlo15}. In the present section I
illustrate an analogous procedure in the framework of the
representation which works for both two- and three-point functions,
with the example of the sunrise diagram with a zero momentum insertion
in one of the propagators (Fig.~1(c)).

Considerable effort has been recently devoted to the investigation of
this diagram.
Asymptotic
expansions have been derived in the papers \cite{adt93,adst93}, and
explicit expressions in terms of generalized hypergeometric, or
Lauricella, functions were obtained in ref.~\cite{inlo17}.
 The same
diagram has also been analyzed in \cite{lunev2}.
It has been noted in \cite{ghin} that a general two-loop diagram can
formally be expressed as a sunrise diagram with masses and the
external momentum being functions of Feynman parameters over which one
can integrate numerically.  The latter reference gave a very convenient
one-dimensional integral representation for this diagram.

The value of the sunrise diagram is
\be
S(p^2,m_1,m_2,m_3)=(\pi e^{\gamma_E})^{2\omega}
\int \!\! \int d^Dk d^Dl {1\over P_1^2 P_2 P_3}
\ee
with $\gamma_E$ being Euler's constant and
\be
P_1&=& (l+k+p)^2-m_1^2+i\eta, \nonumber \\
P_2&=& l^2-m_2^2+i\eta, \nonumber \\
P_3&=& k^2-m_3^2+i\eta,
\ee
and since the sunrise diagram is ultraviolet
divergent we  have to compute it in $D\equiv 4-2\omega$ dimensions.

It has been shown in the previous sections that the triangular regions
over which one has to perform the final two integrations numerically
are determined only by the values of external momenta, and are
independent of masses of particles inside the diagram.  Therefore it
is convenient to choose for the subtraction a diagram which differs
from the diagram we are interested in only by the values of internal
masses.  In the present case we choose a diagram with vanishing $m_2$
and $m_3$ which can be computed analytically
\be
S(p^2,m_1,0,0)&=&-\pi^4\left[
{1\over 2\omega^2}+{1\over 2\omega}(1-2\ln m_1^2) -{1\over 2}
+ {\pi^2\over 4}
\right. \nonumber\\
&& \left. +\ln m_1^2 \left( \ln m_1^2-1 \right)
+{\rm Li}_2 \left(p^2\over m_1^2\right)
 +{p^2-m_1^2\over p^2}\ln\left( {m_1^2-p^2\over m_1^2}
\right)\right]+O(\omega)\nonumber \\ &&
\ee
and the value of a diagram with arbitrary masses can be expressed by
\be
S(p^2,m_1,m_2,m_3)=S(p^2,m_1,0,0)+\Delta(p^2,m_1,m_2,m_3)
\ee
where $\Delta(p^2,m_1,m_2,m_3)\equiv\Delta$
is free from divergences and can be computed using our
method. For simplicity we only consider the case of $p^2<m_1^2$ where
both diagrams are real.

 After the integration over angular variables as in
(\ref{angul}) and over $k_-$ and $l_-$ with help of contour integrals
we obtain
\be
\Delta=
\pi^4 \int \!\! \int_{\cal T} {dk_+ dl_+\over k_+l_+} \int \!\! \int_0^\infty
 ds dt \left(
{A\over (A^2-B^2)^{3/2}}-
{A_0\over (A_0^2-B^2)^{3/2}}\right),
\ee
with
\be
A&=&(2l_++2k_++p)\left( {m_2^2+s\over 2l_+}+{m_3^2+t\over
2k_+}+p\right)-m_1^2-s -t+i\eta\nonumber \\
&\equiv& at+b+cs\nonumber \\
B^2&=&4st
\ee
and the subscript $0$ means that we take $m_2=m_3=0$.
The region of $k_-$ and $l_-$ integration is a triangle ${\cal T}$
delimited by the lines $k_+=0$, $l_+=0$ and $l_++k_+=-p/2$.
The integrations over $s$ and $t$ are easy
\be
\int_0^\infty{\rm d}s {at+b+cs\over \left[(at+b+cs)^2-4st\right]^{3/2}}=
{1\over (1-ac)t-bc}\qquad \mbox{for $a,b,c < 0$}
\ee
and finally we arrive at
\be
\Delta&=&-4\pi^4\int_{-p/2}^0 dk_+\int_{-k_+}^0 {dl_+\over p(2l_++2k_++p)}
\ln {p(2l_++2k_++p)-m_1^2\over
\left(p+{m_2^2\over 2l_+}+{m_3^2\over 2k_+}\right)
(2l_++2k_++p)-m_1^2}.\nonumber\\ &&
\ee
Thus we have found a double integral representation of the sunrise
diagram.  One of the $k_+$, $l_+$ integrations can still be carried
out, and since the argument of the logarithm is a polynomial of the
second degree the result will in general involve dilogarithms of
complex arguments even below the threshold. For the purpose of
numerical evaluation it may be convenient to work with a
double-integral, but explicitly real
representation.
\\
\vspace{1cm}

\noindent
{\bf ACKNOWLEDGMENTS}\\

I thank D.~Broadhurst, K.G.~Chetyrkin, and  A.I.~Davydychev
for discussion and
advice, and B.~Krause and M.~Steinhauser for checking some of
the formulas.  I am very grateful to the organizers of the Carg\`ese
Summer Institute for the opportunity to take part in this great
event. I thank Graduierten\-kol\-leg Elementar\-teilchen\-physik at
the University of Karlsruhe for support.


\begin{thebibliography}{10}

\bibitem{vertexd}
D.~Kreimer,  Phys. Lett. {\bf B292} (1992) 341.

\bibitem{ckk}
A.~Czarnecki, U.~Kilian, and D.~Kreimer,
{\em New representation of two-loop propagator and vertex functions},
\newblock hep-ph/9405423, in press in Nucl.~Phys.~B.

\bibitem{shnote}
D.~Kreimer, {\em
\newblock A short note on two-loop box functions},
\newblock hep-ph/9407234.

\bibitem{rosner67}
J.L.~Rosner, Ann. Phys. {\bf 44} (1967) 11.

\bibitem{che81}
K.G. Chetyrkin and F.V. Tkachov, Nucl. Phys. {\bf B192} (1981) 159.

\bibitem{udyad}
N.I. Ussyukina and A.I. Davydychev,
Phys. Lett. {\bf B298} (1993) 363; Yad. Fiz. {\bf 56} (1993) 172;
Phys. Lett. {\bf B332} (1994) 159.

\bibitem{propagd}
D. Kreimer, Phys. Lett. {\bf B273} (1992) 277.

\bibitem{inlo15}
F.A. Berends and J.B. Tausk, Nulc. Phys. {\bf B421} (1994) 456.

\bibitem{adt93}
A.I. Davydychev and J.B. Tausk,
\newblock Nucl. Phys.  {\bf B397}  (1993) 123.

\bibitem{adst93}
A.I. Davydychev, V.A. Smirnov, and J.B. Tausk,
\newblock Nucl. Phys. {\bf B410} (1993) 325.

\bibitem{inlo17}
F.A. Berends, M.~Buza, M.~B{\"o}hm, and R.~Scharf,
\newblock  Zeit. Phys. {\bf C63} (1994) 227.\\
S. Bauberger, F.A. Berends, M. B\"ohm, and M. Buza, hep-ph/9409388.


\bibitem{lunev2}
F.A. Lunev,
\newblock {\em On evaluation of two-loop self-energy diagram with
three propagators}, \newblock hep-th/9408161.

\bibitem{ghin}
A.~Ghinculov and J.J. {van der Bij},
\newblock {\em Massive two-loop diagrams: the Higgs propagator},
\newblock hep-ph/9405418.

\end{thebibliography}

\end{document}